\documentclass[12pt]{article}
\def\ypiz_2gg{$\pi^0 \rightarrow \gamma\gamma$}
\newcommand {\charex} {$K^+\mathrm{Xe} \rightarrow K^0 p \mathrm{Xe}'$}
\newcommand {\under} {$K^+n \rightarrow K^0 p$}
\newcommand {\dimass} {$m(pK^0_S)$}
\newcommand {\ctb} {$\cos\Theta_K^\mathrm{cm}$}
\newcommand {\cutctb} {$|\cos\Theta_K^\mathrm{cm}| < 0.6$}

\newcommand {\cutcospair} {$\cos\Theta_{pK} > 0.6$}
\newcommand {\pbeam} {$p_\mathrm{beam}$}
\newcommand {\cutpbeam} {$445 < p_\mathrm{beam} < 525$ MeV/c}
\newcommand {\cutpolakao} {$\Theta_p < 100^0$}
\newcommand {\cutpolapro} {$\Theta_K < 100^0$}
\newcommand {\cutrelazi} {$\Phi > 90^0$}

\newcommand {\mtar} {$m^\mathrm{eff}_\mathrm{targ}$}
\newcommand {\ptar} {$p^\mathrm{eff}_\mathrm{targ}$}
\newcommand {\cutmtar} {$m^\mathrm{eff}_\mathrm{targ} > 750$ MeV/c$^2$}

\begin {document}
\title
{Further evidence for formation of a narrow baryon resonance with
positive strangeness in $K^+$ collisions with Xe nuclei}
\author{
DIANA Collaboration\\
V.V. Barmin$^a$, 
A.E. Asratyan$^a$,
V.S. Borisov$^a$, 
C. Curceanu$^b$, \\
G.V. Davidenko$^a$, 
A.G. Dolgolenko$^{a,}$\thanks{Corresponding author. E-mail address:
dolgolenko@itep.ru.},
C. Guaraldo$^b$, 
M.A. Kubantsev$^{a,c}$, \\
I.F. Larin$^a$, 
V.A. Matveev$^a$, 
V.A. Shebanov$^a$, 
N.N. Shishov$^a$, \\
L.I. Sokolov$^a$,
and G.K. Tumanov$^a$\\
\normalsize {$^a$ \it Institute of Theoretical and Experimental Physics,
Moscow 117259, Russia}\\
\normalsize {$^b$ \it Laboratori Nazionali di Frascati dell' INFN,
C.P. 13-I-00044 Frascati, Italy}\\
\normalsize {$^c$ \it Department of Physics and Astronomy,
Northwestern University, Evanston, IL60208, USA}
}                                          
\maketitle

\begin{abstract}
We have continued our investigation of the charge-exchange reaction
\charex\ in the bubble chamber DIANA. In agreement with our previous
results based on part of the present statistics, formation of
a narrow $pK^0$ resonance with mass of $1537 \pm 2$ MeV/c$^2$ is 
observed in the elementary transition \under\ on a neutron bound in 
the Xenon nucleus. Visible width of the peak is consistent with being 
entirely due to instrumental resolution and allows to place an upper 
limit on its intrinsic width: $\Gamma < 9$ MeV/c$^2$. A more precise
estimate of the resonance intrinsic width,
$\Gamma = 0.36\pm0.11$ MeV/c$^2$, is obtained
from the ratio between the numbers of resonant and non-resonant 
charge-exchange events. The signal is observed in
a restricted interval of incident $K^+$ momentum, that is consistent
with smearing of a narrow $pK^0$ resonance by Fermi motion of the 
target neutron. Statistical significance of the signal is some 7.3, 
5.3, and 4.3 standard deviations for the estimators $S / \sqrt{B}$, 
$S / \sqrt{S+B}$, and $S / \sqrt{S + 2B}$, respectively. 
This observation confirms and reinforces our earlier results, and 
offers strong evidence for formation of a pentaquark baryon with 
positive strangeness in the charge-exchange reaction \under\ on a 
bound neutron.
\end{abstract}


\section{Introduction}
     The interest in multiquark baryon states was revived by D. Diakonov, 
V. Petrov, and M. Polyakov \cite{DiPePo} who were able to derive fairly
definite predictions for the low mass and unusually 
small width ($\Gamma \leq 15$ MeV/c$^2$) of the 
strange pentaquark with $J^P = 1/2^+$ and $S = 1$, the $\Theta^+(1530)$. 
They were also able to convince the experimentalists of LEPS at SPring-8 
and DIANA at ITEP Moscow to test these predictions in low-energy
photoproduction and $K^+$--nucleus collisions, respectively. The
enhancements near 1540 MeV/c$^2$ in the $nK^+$ and $pK^0$ effective-mass
spectra, that were almost simultaneously detected by LEPS \cite{LEPS} 
and DIANA \cite{DIANA}, touched off a wave of experimental searches
for the $\Theta^+$ baryon that yielded positive observations as well as
null results (the rapidly evolving experimental situation is described 
in, {\it e.g.,} \cite{review}). Recently, the CLAS experiment reported
null results of high-statistics searches for the $\Theta^+$ in 
photoproduction on hydrogen \cite{CLAS-hydrogen} and deuterium 
\cite{CLAS-deuterium}. In the BELLE experiment at the KEKB asymmetric
$e^+e^-$ collider, interactions of secondary particles with detector
material were used for studying the charge-exchange reaction \under\
in the $K^+$ momentum region below some 2 GeV/c. No signal from 
formation of the $\Theta^+$ baryon was observed, and an upper limit
on the $\Theta^+$ intrinsic width was imposed: $\Gamma < 0.64$ MeV/c$^2$
for $m(\Theta^+) = 1539$ MeV/c$^2$ \cite{BELLE}.
On the other hand, the LEPS and SVD-2 
experiments were able to confirm their earlier positive observations 
at higher statistical levels in $\gamma\mathrm{D}$ collisions with
$E_\gamma \leq 2.4$ GeV \cite{Nakano} and in $pA$ collisions at
$E_p = 70$ GeV \cite{SVD}, respectively.

     The charge-exchange reaction \under\
on a bound neutron, that is investigated by DIANA \cite{DIANA} and 
BELLE \cite{BELLE}, is particularly interesting because it allows to 
probe the $\Theta^+$ intrinsic width in a model-independent manner 
\cite{Strakovsky, Gibbs, Cahn, Sibirtsev-width}. 
Yet another important advantage
of the latter process is that the strangeness of the final-state
$pK^0_S$ system is {\it a priori} known to be positive. In this paper, 
we update the results reported in \cite{DIANA} by analyzing a bigger 
sample of the charge-exchange reaction \under\ in low-energy $K^+$Xe 
collisions.

\section{The experiment and the data}
     The bubble chamber DIANA filled with liquid Xenon was
exposed to a separated $K^+$ beam with momentum of 850 MeV/c from the
10-GeV ITEP proton synchrotron. The density and radiation length of the 
fill are 2.2 g/cm$^3$ and 3.7 cm, respectively. The chamber has a total 
volume of $70\times70\times140$ cm$^3$ viewed by photographic cameras,
and operates without magnetic field \cite{Detector}. Charged particles 
are identified by ionization and momentum-analyzed by range in Xenon.
In the fiducial volume of the bubble chamber, $K^+$ momentum is a
function of longitudinal coordinate and varies from 750 MeV/c for
entering kaons to zero for those that range out through ionization.
(A 150-mm-thick layer of Xenon downstream of the front wall is 
excluded from the fiducial volume, and is only used for detecting the 
secondaries that travel in the backward hemisphere.) Throughout this
interval of $K^+$ momentum, partial cross sections for formation of
various final states of $K^+$Xe collisions can be measured thanks to 
efficient detection of both the decays and interactions of incident
kaons in the Xenon bubble chamber. The momentum of an interacting $K^+$
is determined from longitudinal position of the interaction vertex
with respect to central position of the observed maximum due to decays
of stopping $K^+$ mesons. The uncertainty on $K^+$ momentum, \pbeam,
decreases with increasing \pbeam\ and is near
20 MeV/c for the momentum interval of 450--550 MeV/c. The estimate
of $K^+$ momentum based on measured position of the interaction vertex
has been verified by detecting and reconstructing the 
$K^+ \rightarrow \pi^+\pi^+\pi^-$ decays in flight, which 
provided an independent estimate of $K^+$ momentum.

     On total, some $10^6$ tracks of incident $K^+$  mesons are 
recorded on film. Scanning of the film yielded  nearly 35 000
events with visible $K^0$ decays, $K^0_S \rightarrow \pi^+\pi^-$ and 
$K^0_S \rightarrow \pi^0\pi^0$, that could be associated with primary
$K^+$Xe vertices with different multiplicities of secondary particles.
For each event with a $K^0$ candidate, longitudinal coordinate of the
primary vertex was measured at the scanning stage. Finally, events with 
a single proton and $K^0_S \rightarrow \pi^+\pi^-$ in the final state 
were selected as candidates for the charge-exchange reaction \under\
without rescatterings. In order to reduce the total volume of 
measurements, a lower cutoff was imposed on the $K^+$ track length
before interaction (see further). The selected events are then fully 
measured and reconstructed in space using specially designed 
stereo-projectors similar to those proposed in \cite{cronin}. In 
particular, we measure the $K^0_S$ and proton emission angles with 
respect to the $K^+$ direction, $\pi^+$ and $\pi^-$ emission angles 
with respect to the parent $K^0_S$ direction, and proton and pion 
paths in Xenon. 

     The momentum is estimated by range for 
the proton, and by pion ranges and emission angles for the $K^0_S$. 
In order to reduce the uncertainties on measured momenta and emission 
angles, protons with $p_p < 180$ MeV/c and $K^0_S$ mesons with 
$p_K < 170$ MeV/c are dropped, and the distance between the primary 
vertex and the $K^0_S$ decay vertex is required to exceed 2.5 mm. On 
average, experimental 
resolution is near 2\% for the $K^0$ and proton momenta, and $\sim 2^0$ 
for the angle between their directions in the lab system. Further 
details on the experimental procedure can be found in 
\cite{barmin2, barmin3}. The quality of the data is best reflected 
by experimental resolution on effective mass of the $pK^0$ system, 
estimated as $\sigma_m \simeq 3.0$ MeV/c$^2$ from the simulation. The
latter estimate is consistent with the width of the 
$\Lambda \rightarrow p\pi^-$ peak measured in a previous experiment
with the same detector \cite{Lambda,DIANA}.

\begin{figure}[h]

\vspace{6.5cm}
\includegraphics{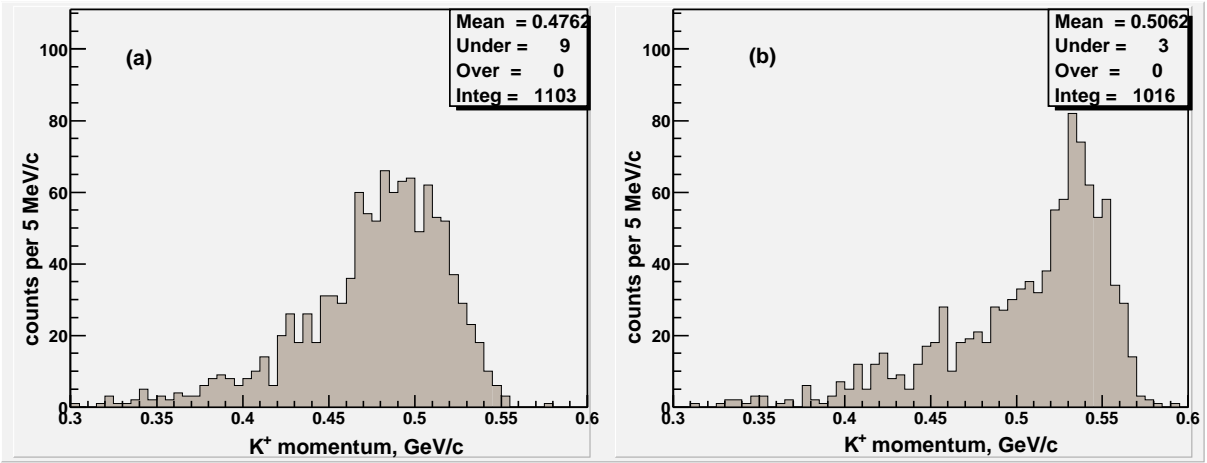}
\caption
{Incident $K^+$ momentum for measured events of the reaction \charex.
The old and new data are shown in (a) and (b), respectively.}
\label{pbeam}
\end{figure}

     This work is based on a sample of 2131 fully measured events of 
the charge-exchange reaction \charex, that comprises the ``old" data
analyzed in \cite{DIANA} as well as the ``new" data analyzed here for
the first time. The $K^+$ range before interaction was required to exceed 
550 mm for the ``old" data, and 520 mm for the 
``new" data. On average, this corresponds to $K^+$ momentum cutoffs of 
\pbeam\ $< 530$ and 560 MeV/c, respectively (the correspondence is not 
exact since beam momentum varied by some $\pm 20$ MeV/c in different 
exposures). For all measured events of the  reaction \charex, momentum 
spectra of incident kaons are separately plotted for the old and new 
data in Fig. \ref{pbeam}. Note that compared to our
earlier analysis \cite{DIANA}, the statistics of the charge-exchange
reaction has nearly doubled, and mean $K^+$ momentum has increased
from 470 to 500 MeV/c. Expanding the range of $K^+$ momentum allows to
probe the lineshape of the putative $pK^0$ resonance that is smeared
by Fermi motion of the target nucleon, see further. 

\section{The simulation and the effects of rescattering}
     Rescattering of either the $K^0$ or proton in the Xenon nucleus
distorts the effective mass of the $pK^0$ system originally formed
in the charge-exchange reaction \under. In order to suppress the
background arising from rescatterings, the following selections were 
used in our previous analysis \cite{DIANA}: \cutpolakao, \cutpolapro,
and \cutrelazi. Here, $\Theta_K$ and $\Theta_p$ are the $K^0$ and
proton emission angles with respect to the $K^+$ direction in the
lab system, and $\Phi$ is the angle between the $K^0$ and proton in
the plane normal to beam direction. The validity of these selections
was verified by the theoretical analysis \cite{Sibirtsev}.

     We demonstrate the rejection power of the above selections 
to rescatterings using a simple Monte-Carlo simulation of the 
charge-exchange reaction \under\ in nuclear environment. In the 
simulation, total energy of a bound neutron is parametrized as
$E_n = m_N - 2\epsilon - p_n^2 / (2 m_N)$, where $m_N$ is the mass of 
a free neutron, $\epsilon$ is the mean binding energy, and $p_n$ is 
Fermi momentum \cite{Sibirtsev}. For the Xenon nucleus, we assume
$\epsilon = 7$ MeV and use a realistic form of the Fermi-momentum
distribution with maximum near 170 MeV/c. The flux of
incident $K^+$ mesons as a function of $K^+$ momentum is inferred
from the observed distribution of $K^+$ range in Xenon before 
interaction or decay, see \cite{DIANA}. The effects of
apparatus resolution and measurement errors are included in the
simulation. Rescattering in the nucleus is not accounted for. 
In the distributions to follow, the number of simulated events is 
normalized to that of observed ones prior to cuts.
\begin{figure}[t]

\vspace{6.5cm}
\includegraphics{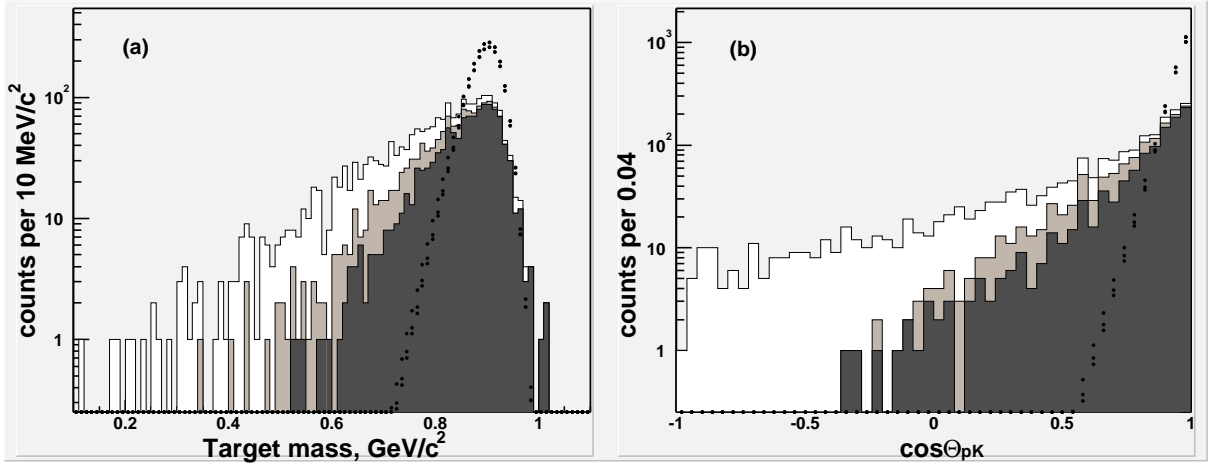}
\caption
{The mass of effective target (a) and the angle between the $pK^0$
and $K^+$ directions in the lab system (b) for observed and simulated 
events (simulated distributions are depicted by dots).
The middle histograms (light-shaded and dotted) are 
for the selections \cutpolakao\ and \cutpolapro\ only. The bottom
histograms (dark-shaded and dotted) are for the full selections
$\Theta_K, \Theta_p < 100^0$ and \cutrelazi.}
\label{mtar-and-cospair}
\end{figure}

     We find that the selections \cutpolakao, \cutpolapro, and 
\cutrelazi\ reject some 12\% of the simulated rescattering-free
events ($\sim 11$\% without \cutrelazi), and as much as $\sim 46$\% 
of all live events of the reaction \under\ detected in our 
experiment. This shows that the above selections
significantly reduce the fraction of rescattered events in the sample.
For the binary reaction considered, it is straightforward to estimate 
the mass of the effective target, \mtar. The \mtar\ distributions of
live and simulated events prior to cuts, for the selections \cutpolakao\
and \cutpolapro\ only, and for the full selections 
$\Theta_K, \Theta_p < 100^0$ and \cutrelazi\ are shown in 
Fig. \ref{mtar-and-cospair}a. The observed \mtar\ 
distribution reaches a maximum near that of the simulated distribution,
but shows a long downward tail that is due to rescatterings, see 
Fig. \ref{mtar-and-cospair}a. The aforementioned selections 
are seen to substantially reduce the width of the \mtar\ distribution 
by rejecting rescatterings. Cosine of the angle between the $pK^0$ and 
incident $K^+$ directions in the lab frame, $\Theta_{pK}$, is plotted
in Fig. \ref{mtar-and-cospair}b. Due to rescatterings, live $pK^0$ 
systems are typically emitted at broader angles to the $K^+$ beam than 
the simulated ones. Again, the aforementioned angular selections lead
to a better agreement between the observed and simulated distributions.
\begin{figure}[ht]

\vspace{6.5cm}
\includegraphics{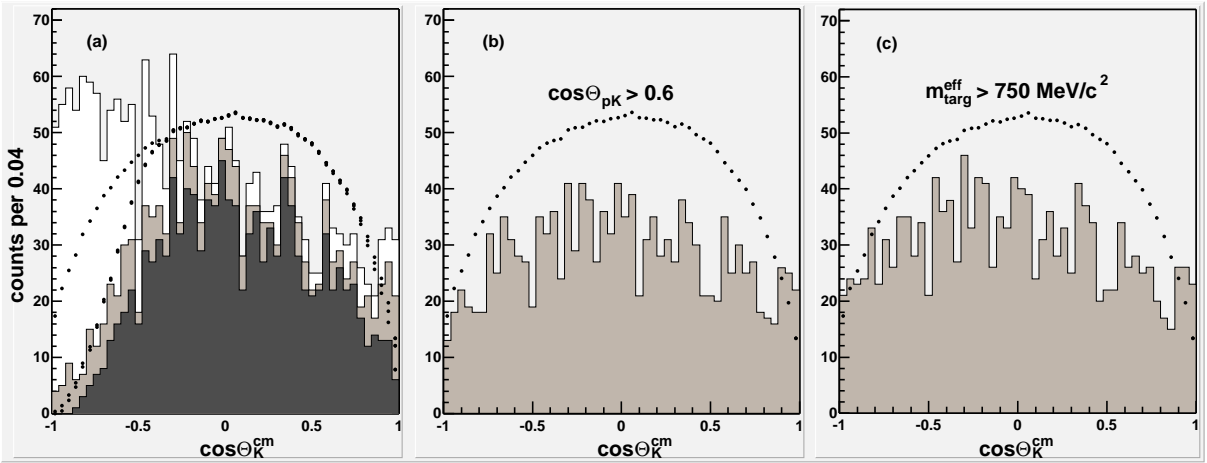}
\caption
{Cosine of the $K^0$--$K^+$ angle in the $pK^0$ rest frame,  
$\Theta_K^\mathrm{cm}$, for observed and simulated events (simulated
distributions are depicted by dots).
In (a), the middle histograms (light-shaded and dotted) are for
the selections \cutpolakao\ and \cutpolapro\ only, and the bottom
histograms (dark-shaded and dotted) are for the full selections
$\Theta_K, \Theta_p < 100^0$ and \cutrelazi.
The alternative selections 
\cutcospair\ and \cutmtar\ result in the observed and simulated 
\ctb\ distributions shown in (b) and (c), respectively.}
\label{cm-cosine}
\end{figure}
It is interesting to see how the rescatterings affect the angular
distribution in the $pK^0$ center-of-mass frame. Cosine of the 
$K^0$--$K^+$ angle in the $pK^0$ rest frame, 
$\Theta_K^\mathrm{cm}$,
is shown in Fig. \ref{cm-cosine}a. Prior to selections, the \ctb\
distribution of live events monotonically decreases between -1 and
+1, whereas that of simulated events is almost forward--backward
symmetric (the dips towards the boundary values are due to lower cuts
on the $K^0$ and proton momenta, see above). The selections 
\cutpolakao, \cutpolapro, and \cutrelazi\ are seen to restore the
agreement between the simulated and observed \ctb\ distributions 
by suppressing rescatterings. Alternatively, the (approximate) 
forward--backward symmetry of the \ctb\ distribution is restored by
applying other selections suggested by the simulation: \cutcospair\
or \cutmtar\ (see Figs. \ref{cm-cosine}b and \ref{cm-cosine}c).

\section{The $pK^0$ effective-mass distributions}
\begin{figure}[!t]

\vspace{13cm}
\includegraphics{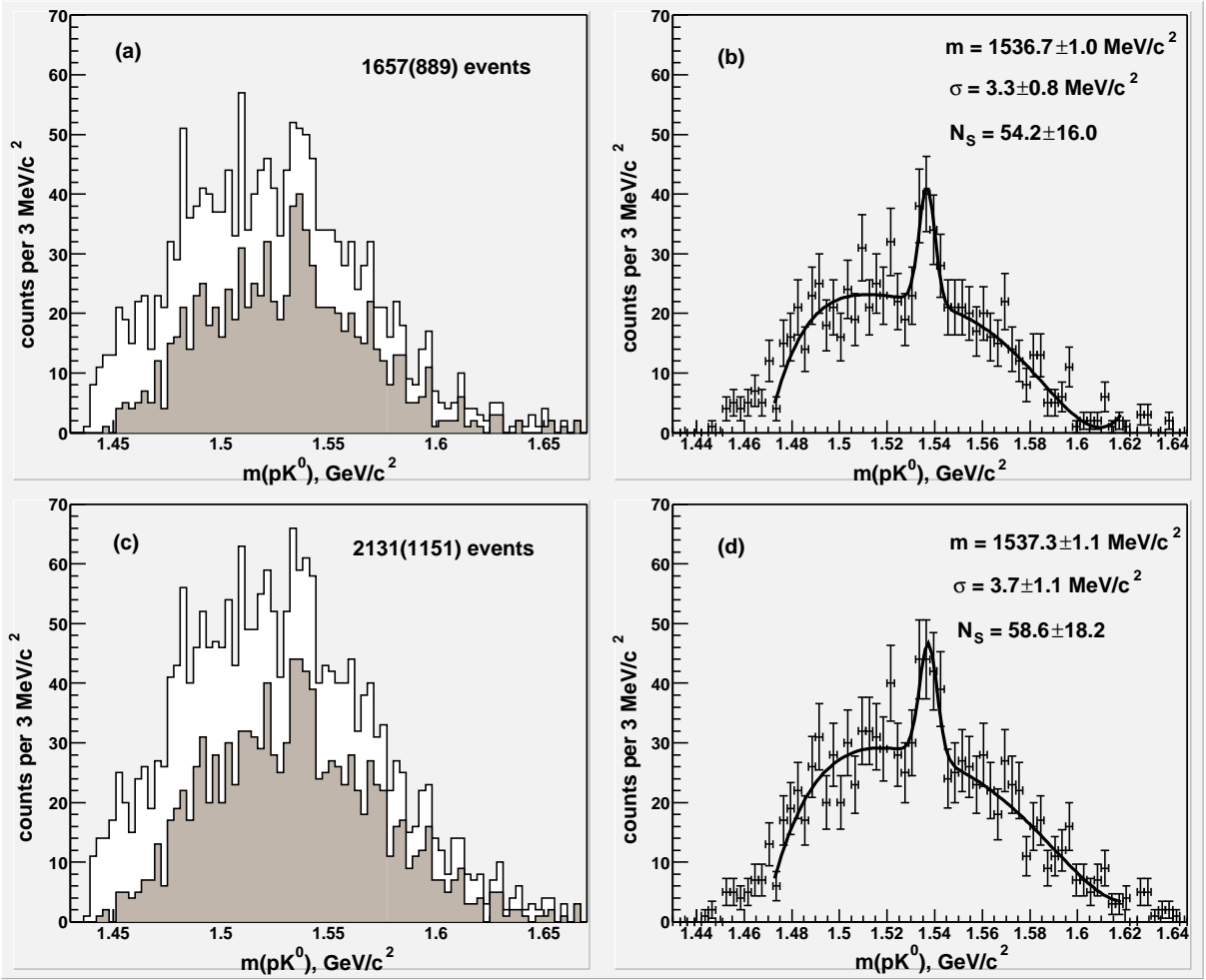}
\caption
{Effective mass of the $pK^0$ system formed in the reaction 
\charex\ for \pbeam\ $< 530$ MeV (a) and for the full range of
\pbeam\ (c). Shaded histograms result from the selections 
\cutpolakao, \cutpolapro, and \cutrelazi. In (b) and (d),
corresponding shaded histograms are fitted to a Gaussian on
top of a fifth-order polynomial.}
\label{as-before}
\end{figure}

     The initial stage of this analysis is to approximate the 
experimental conditions of \cite{DIANA}, and to use exactly the same
selections as in \cite{DIANA}, in order to see what happens with the 
putative signal. So here we apply the selection \pbeam\ $< 530$ MeV/c
that corresponds to the effective \pbeam\ cutoff for the old data. 
Effective mass of the $K^0 p$ system formed in the charge-exchange 
reaction, \dimass, is shown in Fig. \ref{as-before}a.
(Note that because of the cut \pbeam\ $< 530$ MeV/c, relative 
contribution of the "new" data is only $\simeq 37$\% in 
Fig. \ref{as-before}a.) Prior to selections aimed at suppressing
the background from proton and $K^0$ rescattering in the Xenon 
nucleus, but a small enhancement is seen in the mass region of
1535--1540 MeV/c$^2$. The selections \cutpolakao, \cutpolapro, and 
\cutrelazi\ reduce the $pK^0$ effective-mass spectrum to the shaded 
histogram shown in the same Figure. These selections are seen to 
emphasize the enhancement in the mass region of 1535--1540 MeV/c$^2$. 
The latter \dimass\ distribution is then fitted to a Gaussian
on top of a fifth-order polynomial, see Fig. \ref{as-before}b. The
width of the observed peak is compatible with experimental resolution
on \dimass. Compared to \cite{DIANA} where a signal of $\sim 30$ events
above background at \dimass\ $= 1539 \pm 2$ MeV/c$^2$ was reported,
we now observe a substantially bigger signal at a slightly smaller 
$pK^0$ effective mass. Analogous \dimass\ distributions
for the full interval of \pbeam\ are shown in Figs. \ref{as-before}c 
and \ref{as-before}d. Adding events with \pbeam\ $> 530$ MeV/c 
results in a broader enhancement and a higher level of background.

\begin{figure}[h]

\vspace{6.5cm}
\includegraphics{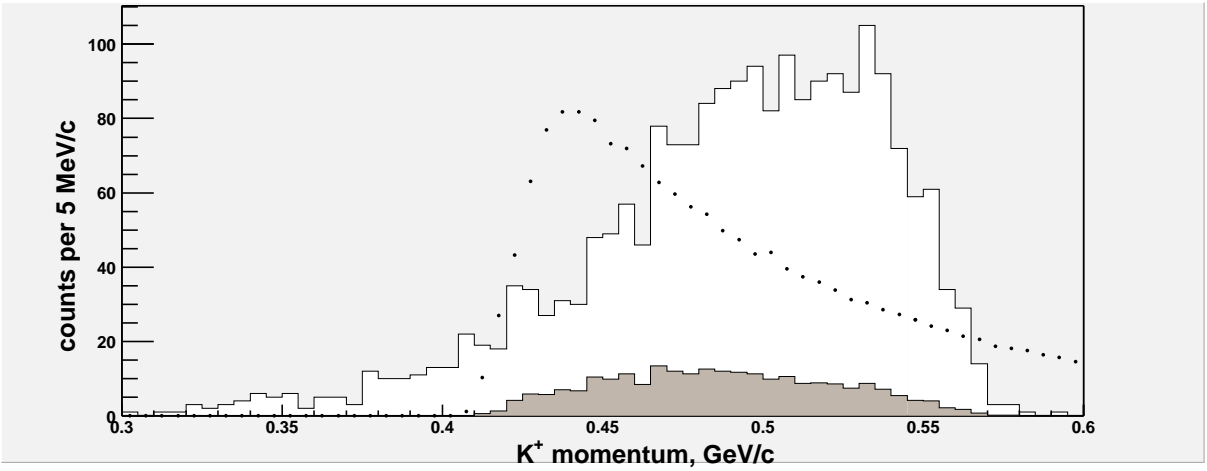}
\caption
{Open histogram: incident $K^+$ momentum for all measured events
of the charge-exchange reaction \under. Dotted histogram:
for an assumed narrow $pK^0$ resonance with mass of 1537 MeV/c$^2$,
the simulated ratio between the resonant and non-resonant contributions
to \under\ as a function of $K^+$ momentum (in arbitrary units). Shaded 
histogram: predicted lineshape of the assumed $pK^0$ resonance. 
For definiteness, the latter distribution has been normalized to the
number of live events in the mass interval $1532 < m(pK^0) < 1544$ 
MeV/c$^2$ prior to selections.}
\label{smart}
\end{figure}

     It is interesting to study the dependence of the putative signal
on incident $K^+$ momentum, \pbeam. Note that a narrow $pK^0$ resonance
would produce a line in the \pbeam\ distribution for \under\ if the
target neutron were free rather than bound, whereas the Fermi motion 
would smear the above line to a lineshape of finite width. On the
Monte-Carlo level, we simulate formation of a narrow $pK^0$ resonance
with mass of 1537 MeV/c$^2$ by selecting events in the \dimass\ interval 
of $1537 \pm 2$ MeV/c$^2$ as the resonant contribution to the 
charge-exchange reaction \under. The simulated ratio between the 
resonant and non-resonant contributions to \under\ as a function of 
$K^+$ momentum, that is independent of the $K^+$ flux, is shown in 
Fig. \ref{smart} (arbitrary units). Despite the smearing, the peak is 
seen to survive in the form of a broad asymmetric maximum near the 
resonant value of \pbeam\ for a free target neutron. Now, multiplying 
the \pbeam\ distribution of all detected charge-exchange events by the 
aforementioned ratio, we are able to estimate the \pbeam\ distribution
for a narrow $pK^0$ resonance with $m = 1537$ MeV/c$^2$ in our 
experimental conditions, see Fig. \ref{smart}. The lineshape of 
the assumed resonance has a smaller width than the \pbeam\
distribution of all \under\ events, which suggests that a true
resonant signal should benefit from restricting the range of $K^+$ 
momentum. 

\begin{figure}[t]

\vspace{13cm}
\includegraphics{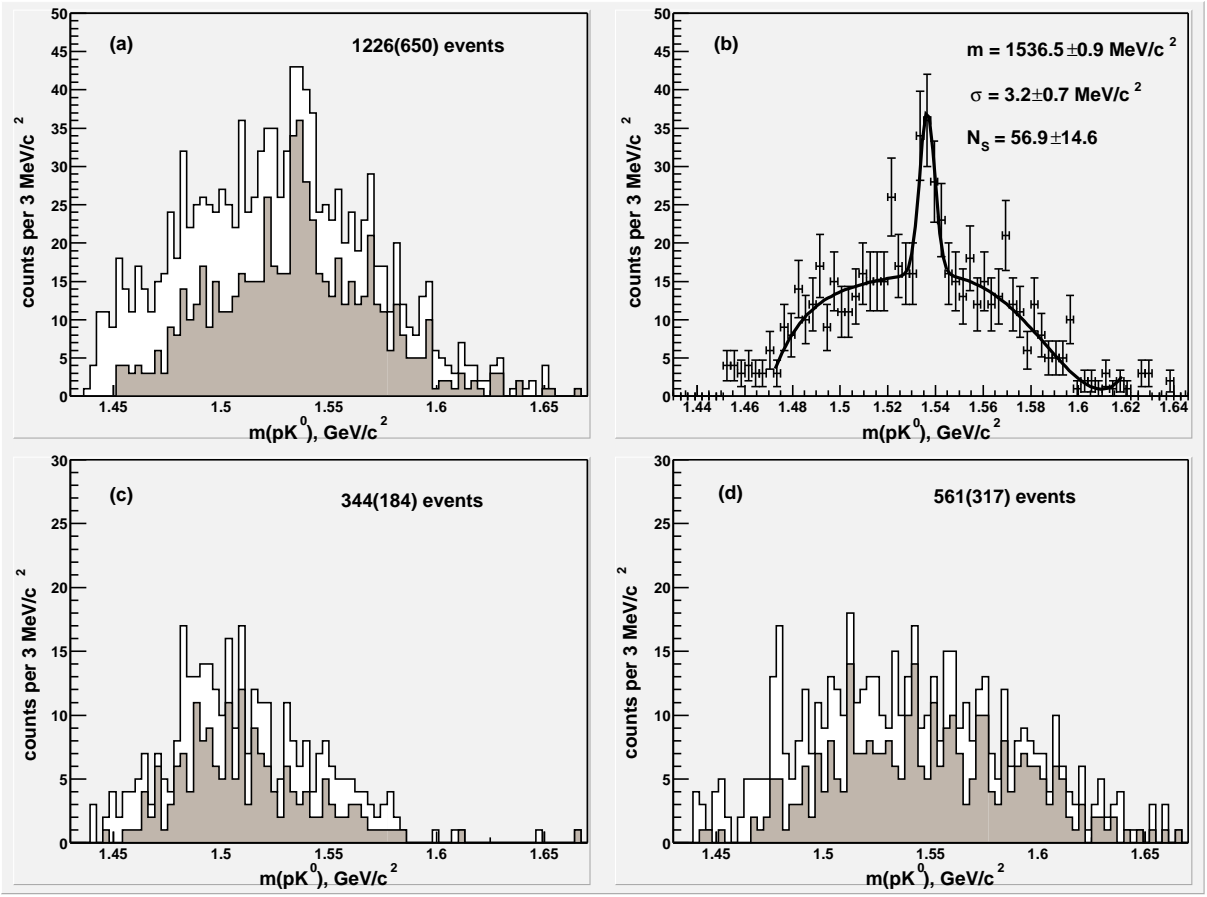}
\caption
{Effective mass of the $K^0 p$ system for \cutpbeam\ (a),
\pbeam\ $< 445$ MeV/c (c), and \pbeam\ $> 525$ MeV/c (d).
Shaded histograms are for the selections \cutpolakao, 
\cutpolapro, and \cutrelazi. Shown in (b) is a fit of the 
shaded distribution in (a) to a Gaussian on top of a 
fifth-order polynomial.}
\label{narrow-and-wings}
\end{figure}

\begin{figure}[ht]

\vspace{13cm}
\includegraphics{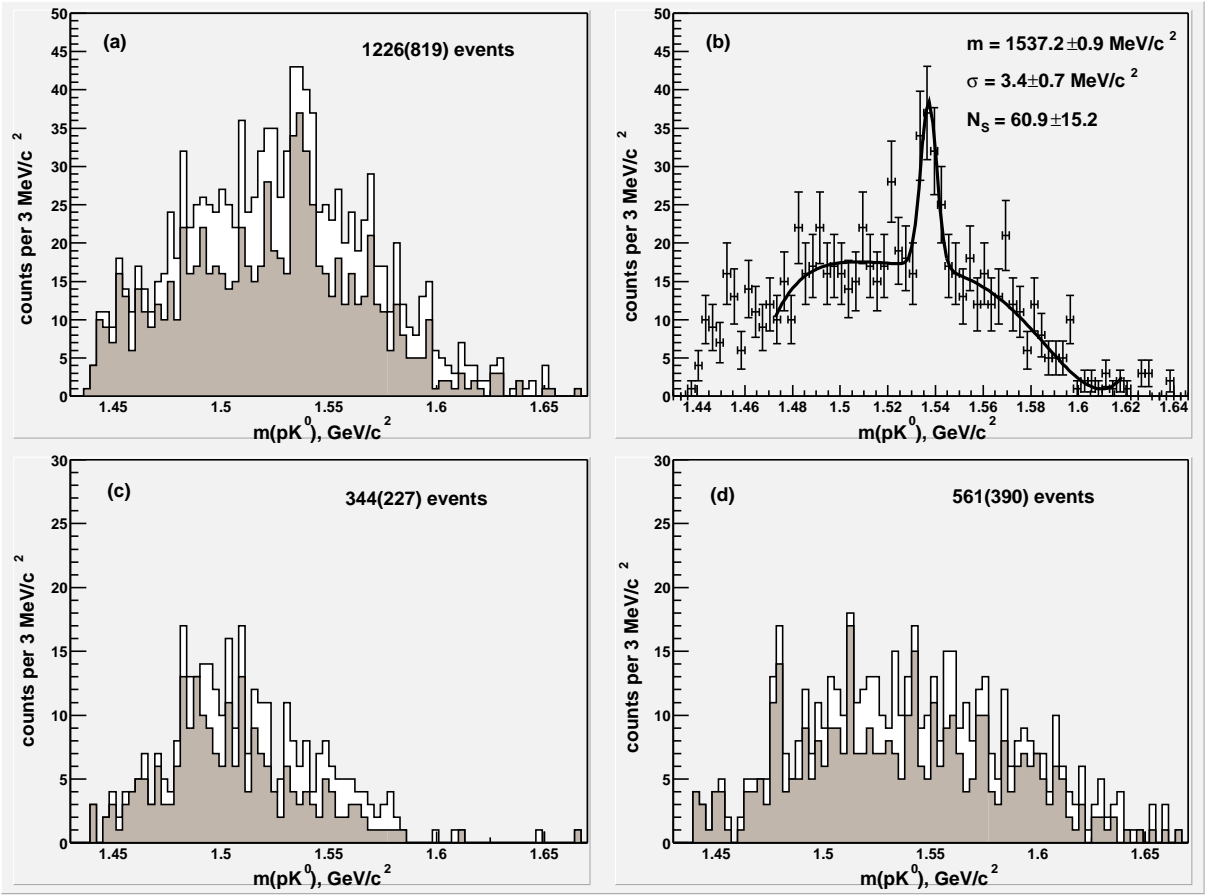}
\caption
{Similar \dimass\ distributions as in Fig. \ref{narrow-and-wings},
but shaded histograms are for the selections \cutpolakao\ and 
\cutpolapro\ only (that is, the selection \cutrelazi\ is lifted).}
\label{noazi}
\end{figure}

     Restricting the $K^+$ momentum to the interval \cutpbeam, that 
is consistent with Fermi-smearing of a narrow resonance, results in
a prominent peak near 1537 MeV/c$^2$ in the $pK^0$ effective-mass 
spectrum, see Fig. \ref{narrow-and-wings}a. Fitted width of the peak is 
again consistent with being entirely due to apparatus smearing of the 
$pK^0$ effective mass. The \dimass\ distributions for the momentum 
intervals \pbeam\ $< 445$ MeV/c and \pbeam\ $> 525$ MeV/c are 
structureless, see Figs. \ref{narrow-and-wings}c and 
\ref{narrow-and-wings}d. The latter is in contradiction with the
claim \cite{Zavertyaev} that the $pK^0$ peak reported in \cite{DIANA}
is a kinematic artifact of $K^+n$ collisions on quasifree neutrons.
Similar \dimass\ distributions but for the selections \cutpolakao\
and \cutpolapro\ only, that is, with the cut \cutrelazi\ lifted, are 
shown in Fig. \ref{noazi}. Lifting the selection \cutrelazi\ leaves
the peak near 1537 MeV/c$^2$ virtually unaffected, but broadens the
background under the peak. Statistical
significance of the signal in Fig. \ref{noazi}b is near 7.3, 5.3, and
4.3 standard deviations for the estimators $S / \sqrt{B}$,
$S / \sqrt{S+B}$, and $S / \sqrt{S + 2B}$, respectively. Here, 
$B \simeq 68$ events and $S \simeq 60$ events are the fitted 
background and the excess above background in the mass interval 
$1532 < m(pK^0) < 1544$ MeV/c$^2$ that is consistent with the 
instrumental resolution on \dimass.

\begin{figure}[t]

\vspace{13cm}
\includegraphics{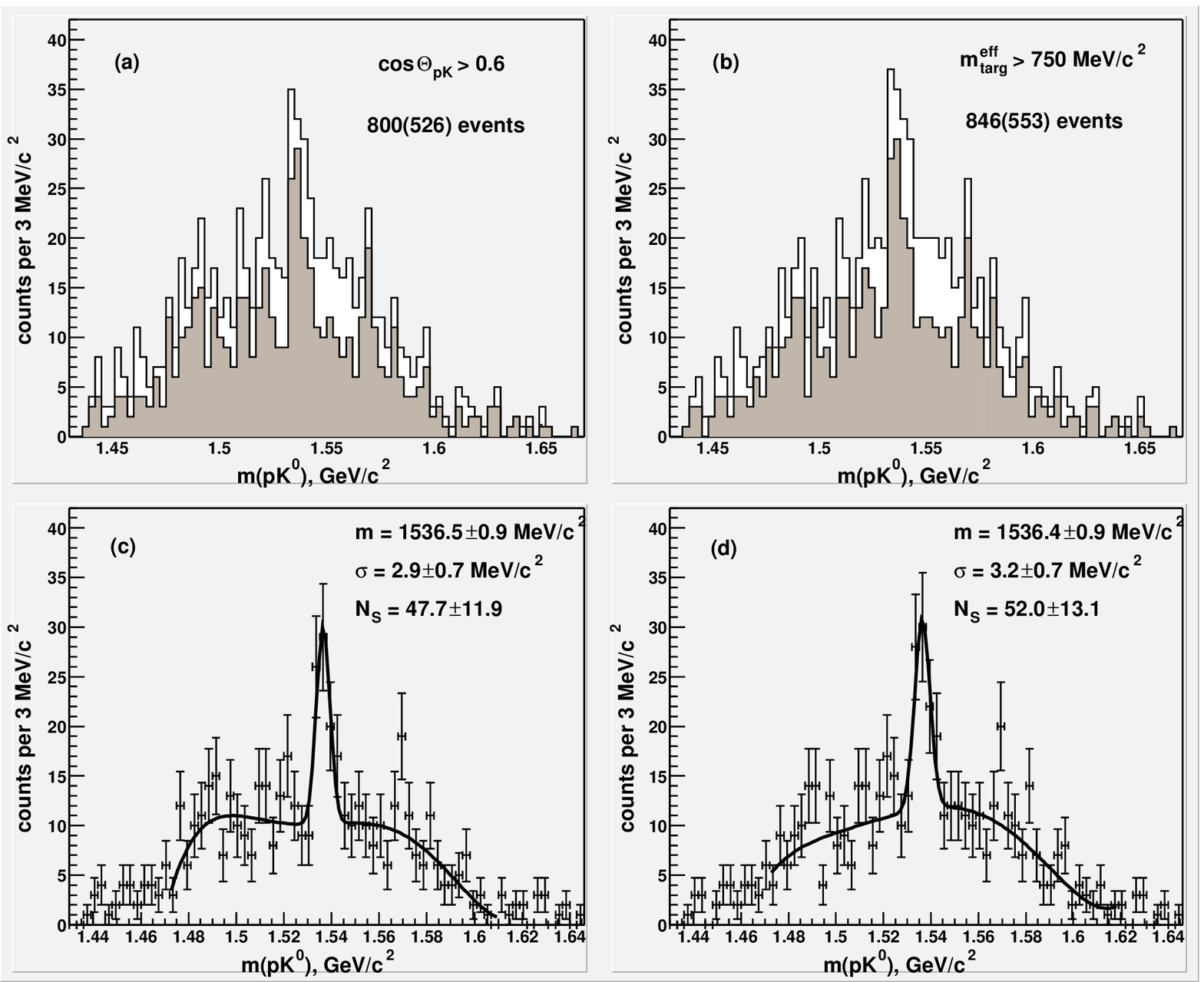}
\caption
{Effective mass of the $K^0 p$ system for \cutpbeam, plotted under
the selections \cutcospair\ (a) or \cutmtar\ (b). Shaded histograms
are for the additional selection \cutctb. Fits of the latter
distributions are shown in (c) and (d).}
\label{new-cuts}
\end{figure}

     The data of Figs. \ref{cm-cosine} and \ref{mtar-and-cospair}
suggest the feasibility of alternative selections. The \dimass\
distributions for \cutpbeam, plotted under either \cutcospair\ or
\cutmtar, are shown in Figs. \ref{new-cuts}a and \ref{new-cuts}b,
respectively. Under either selection, the distribution of the 
center-of-mass angle $\Theta_K^\mathrm{cm}$ is flatter for the live 
than simulated events, see Fig. \ref{cm-cosine}. The latter suggests 
that the background fraction increases towards the edges, 
$\cos\Theta_K^{\mathrm{cm}}= \pm 1$. Therefore, we use an additional
selection \cutctb\ in conjunction with either \cutcospair\ or \cutmtar.
Either combined selection rejects $\sim 30$\% of simulated \under\
events. The resulting $pK^0$ mass spectra for \cutpbeam, that are 
illustrated and fitted in Fig. \ref{new-cuts}, show prominent peaks 
similar to that in Fig. \ref{narrow-and-wings}b. The corresponding 
\dimass\ distributions for the $K^+$ momentum intervals 
\pbeam\ $< 445$ MeV/c and \pbeam\ $> 525$ MeV/c (not shown) 
again prove to be featureless. We find that absolute momentum of the
effective target, \ptar, is strongly correlated with \mtar, so that
the selection \cutmtar\ is roughly equivalent to \ptar\ $< 400$ MeV/c
(corresponding \dimass\ distributions are not shown here).

\section{Intrinsic width of the $\Theta^+$ baryon}
     The width of the observed $pK^0$ peak is consistent with being 
entirely due to instrumental smearing and allows to place an upper limit 
on intrinsic width: $\Gamma < 9$ MeV/c$^2$. On the other hand, intrinsic 
width of a $pK^0$ resonance formed in the charge-exchange
reaction \under\ can be determined as \cite{Cahn}
\begin{displaymath}
\Gamma = \frac{N^\mathrm{peak}}{N^\mathrm{bkgd}}
\times \frac{\sigma^\mathrm{CE}}{107\mathrm{mb}}
\times \frac{\Delta m}{B_i B_f},
\end{displaymath}
where $N^\mathrm{peak}$ and $N^\mathrm{bkgd}$ are numbers of events
in the peak and in the charge-exchange background under the peak,
$\sigma^\mathrm{CE} = 4.1\pm0.3$ mb is the cross section for 
\under\ \cite{cross-section}, $B_i$ and $B_f$
are branching fractions for the initial and final states
($B_i = B_f = 1/2$), and $\Delta m$ is the \dimass\ interval under
the peak that is populated by $N^\mathrm{bkgd}$ background events.
From the signal-to-background ratio observed in our previous analysis
\cite{DIANA}, the $\Theta^+$ width was estimated in \cite{Cahn} as
$\Gamma = 0.9\pm0.3$ MeV/c$^2$. The assumption was made that the
cuts adopted in \cite{DIANA} reduced the resonant and non-resonant
charge-exchange processes by the same factor, and systematic 
uncertainties associated with rescattering in the nucleus could not
be evaluated. But as soon as the $\Theta^+$ decay width is $\sim 1$ 
MeV or less, the bulk of produced $\Theta^+$ baryons will decay
outside of the nucleus giving rise to $pK^0$ pairs that are not 
affected by rescatterings. The latter assumption is indirectly 
supported by the data: we see that expanding the \pbeam\ interval
beyond the resonant region rapidly degrades the signal-to-background
ratio while the $\Theta^+$ signal remains the same within errors.
So for a self-consistent determination of $\Gamma$, 
the ``original" rather than observed number of nonresonant $pK^0$ 
pairs under the $\Theta^+$ peak should be substituted in the above 
formula. 

     We estimate the non-resonant $pK^0$ background not distorted by 
rescatterings from the simulated \dimass\ distribution, applying the
selections $\Theta_K, \Theta_p < 100^0$ and \cutpbeam\ as in 
Fig. \ref{noazi}. In the $K^+$ momentum unterval \cutpbeam, the number
of simulated events prior to experimental selections and angular cuts
is normalized to the original number of all charge-exchange collisions 
with $K^0_S \rightarrow \pi^+\pi^-$ in the final state. The latter is
estimated as $4060\pm400$ collisions from the scanning information. 
Of these events, $(60\pm7)$\% are estimated to survive upon rejecting 
$K^0_S$ mesons with $L < 2.5$ mm, protons and $K^0_S$ mesons that have 
reinteracted in liquid Xenon, and unmeasurable events. Normalizing the
number of simulated events to $2440\pm370$ live events, we then apply 
the cuts $p_K > 170$ MeV/c, $p_p > 180$ MeV/c, \cutpolakao, and 
\cutpolapro\ on the Monte-Carlo level assuming no rescattering in
the nucleus. Thereby, the ``original" non-resonant background in the 
mass region $1532 < m(pK^0) < 1544$ MeV/c$^2$ is 
estimated as $N^\mathrm{bkgd} \simeq 310\pm47$ events. Substituting 
this value in the above formula together with the observed signal for 
the same selections, $N^\mathrm{peak} = 60\pm15$ events, we obtain 
$\Gamma = 0.36\pm0.11$ MeV/c$^2$ where the error does not include
systematic uncertainties of the simulation procedure. This 
experimental estimate has been obtained assuming that the bulk of
produced $\Theta^+$ baryons neither decay nor reinteract inside the
nucleus. Our estimate of the $\Theta^+$ intrinsic width does not 
contradict the upper limit set by BELLE \cite{BELLE}.

\section{Summary and conclusions}
     To summarize, a narrow $pK^0$ resonance has been observed in 
the charge-exchange reaction \under\ on a neutron bound in the Xenon
nucleus. The mass of the resonance is estimated as
$m = 1537 \pm 2$ MeV/c$^2$. Visible width of the peak, 
$\sigma = 3.4 \pm 0.7$ MeV/c$^2$, is consistent with being entirely 
due to instrumental resolution and allows to place an upper limit on 
intrinsic width: $\Gamma < 9$ MeV/c$^2$. 
A more precise estimate of the resonance intrinsic width,
$\Gamma = 0.36\pm0.11$ MeV/c$^2$, has been obtained
from the ratio between the numbers of resonant and non-resonant 
charge-exchange events. The signal is observed in
a restricted interval of incident $K^+$ momentum, that is consistent
with smearing of a narrow $pK^0$ resonance by Fermi motion of the 
target neutron. With an excess of $\sim 60$ events over a fitted
background of $\sim 68$ events, statistical significance of the 
signal is some 7.3, 5.3, and 4.3 standard deviations for the 
estimators $S / \sqrt{B}$, $S / \sqrt{S+B}$, and $S / \sqrt{S + 2B}$, 
respectively. We interpret this 
observation as strong evidence for formation of a pentaquark baryon 
with positive strangeness in the charge-exchange reaction \under\ on 
a bound neutron. The results reported in this paper confirm and 
reinforce our earlier observation based on part of the present 
statistics of the charge-exchange reaction \cite{DIANA}. The
measurements and data analysis are still in progress.

     We wish to thank K. Boreskov, A. Kudryavtsev, T. Nakano, 
I. Strakovsky,\\ G.H.~Trilling, and M. Zhalov for stimulating discussions 
and useful comments. This work is supported by the Russian Foundation 
for Basic Research (grant 04-02-17467).

\end{document}